\documentclass[aps,pre,preprint,amsfonts,amssymb,amsmath,a4paper,10pt]{revtex4-1}
\usepackage[english]{babel}
\usepackage[T1]{fontenc}
\usepackage[latin1]{inputenc}
\usepackage{microtype}
\usepackage{xspace}
\usepackage[caption=false]{subfig}
\usepackage{graphicx}
\usepackage{textcomp}
\usepackage{MnSymbol}
\usepackage{pifont}
\usepackage[absolute]{textpos}
\begin{document}
\newcommand{\euler}[1]{\mathrm{e}^{#1}}
\newcommand{\abschnitt}[1]{section~\ref{#1}\xspace}
\newcommand{\bild}[1]{Fig.~\ref{#1}\xspace}
\newcommand{\gleich}[1]{Eq.~\eqref{#1}\xspace}
%
\newcommand{\mitdimen}[1]{\overset{\smash\sim}{\vphantom{\text{\scriptsize{D}}}\smash #1}\xspace}
\newcommand{\mitdimenflach}[1]{\overset{\smash\sim}{\vphantom{\text{\scriptsize{x}}}\smash #1}\xspace}
\newcommand{\rhomitdimen}{\overset{\,\smash\sim}{\vphantom{\text{\scriptsize{x}}}\smash \rho}\xspace}
\newcommand{\phimitdimen}{\overset{\,\smash\sim}{\vphantom{\text{\scriptsize{D}}}\smash\phi}\xspace}
\newcommand{\ohnedimen}[1]{#1\xspace}
\newcommand{\ohnedimenflach}[1]{#1\xspace}
\newcommand{\ohnedimenflachb}[1]{#1\xspace}
\newcommand{\ohnedimenflachc}[1]{#1\xspace}
\newcommand{\rhoohnedimen}{\rho\xspace}
\newcommand{\phiohnedimen}{\phi\xspace}
\newcommand{\definiert}{\mathrel{\mathop{:}}=}
\newcommand{\definiertb}{=\mathrel{\mathop{:}}}
\newcommand{\partiell}[2]{\frac{\partial #1}{\partial #2}}

\title{Electric-double-layer structure close to the three-phase contact line in an electrolyte wetting a solid substrate}
\date{\today}
\author{Aaron D\"orr}
\author{Steffen Hardt}
\thanks{Email address for correspondence: hardt@csi.tu-darmstadt.de}
\affiliation{Institute for Nano- and Microfluidics, Center of Smart Interfaces, Technische Universit\"at Darmstadt, Petersenstra\ss e 32, 64287 Darmstadt, Germany}
%
%
\begin{textblock*}{21cm}(2.1cm,0.5cm)
\noindent\small \textit{The following article has been published under Phys. Rev. E 86, 022601 (2012) and is available at\\ \mbox{http://pre.aps.org/abstract/PRE/v86/i2/e022601}}
\end{textblock*}
\begin{abstract}
The electric-double-layer structure in an electrolyte close to a solid substrate near the three-phase contact line is approximated by considering the linearized Poisson-Boltzmann equation in a wedge geometry. The mathematical approach complements the semi-analytical solutions reported in the literature by providing easily available characteristic information on the double layer structure. In particular, the model contains a length scale that quantifies the distance from the fluid-fluid interface over which this boundary influences the electric double layer. The analysis is based on an approximation for the equipotential lines. Excellent agreement between the model predictions and numerical results is achieved for a significant range of contact angles. The length scale quantifying the influence of the fluid-fluid interface is proportional to the Debye length and depends on the wall contact angle. It is shown that for contact angles approaching~90$^\circ$ there is a finite range of boundary influence. 
\end{abstract}
\maketitle
The wetting of solid substrates by electrolyte solutions plays an important role in microfluidics, specifically in digital microfluidics based on the electrostatic actuation of sessile droplets~\cite{Fair2007,Abdelgawad2009}. In this context, it is important to study the influence of electric fields around the three-phase contact line. In electrowetting-on-dielectric, it has been shown that the large magnitude of the electric field close to the contact line can be the cause of contact-angle saturation~\cite{Peykov2000,Papathanasiou2005}. Also without applying an external electric field various challenges remain, which have been addressed in Refs.~\cite{Kang2003a,Kang2003b,Chou2001,Monroe2006a} focusing on the electrostatic contribution to wetting. It has been shown that the energetic contribution of the electric double layer close to the three-phase contact region represents a significant fraction of the total free energy of a sessile droplet that is moderately sized compared to the Debye length~\cite{Kang2003a,Kang2003b,Chou2001,Monroe2006a,Monroe2006b}. However, little is known about the structure of the electric double layer in the contact-line region, a topic we address in the present study. Hence we consider the three-phase contact region of two immiscible fluids, e.g.~water (index $w$) and oil (index $np$ for nonpolar), and a solid substrate (index $s$); see \bild{fig:Physical_problem}.
\begin{figure}[ht]%
\centering%
\subfloat[]{\includegraphics{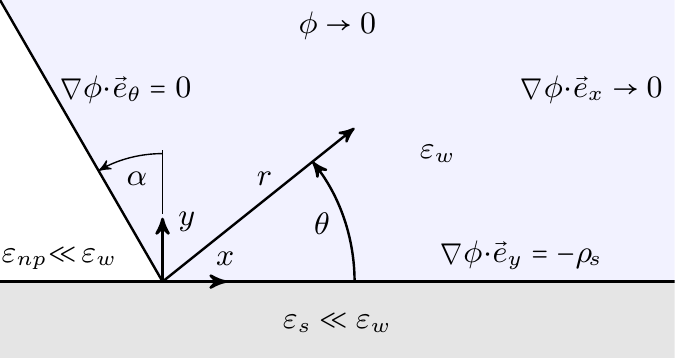}\label{fig:Physical_problem}}\hspace{5ex}
\subfloat[]{\includegraphics{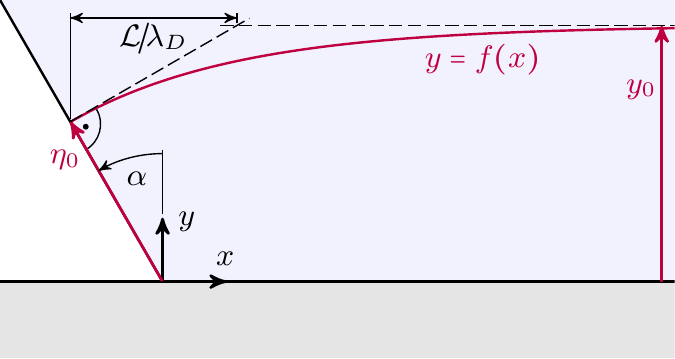}\label{fig:Variables}}
\caption[]{ Schematics of the physical problem and the approximation: \subref{fig:Physical_problem} geometry, material properties, and boundary conditions for the potential~$\phiohnedimen$ and \subref{fig:Variables} principle of approximation, parametrization of the functional form of the equipotential lines, and definition of the length scale~$\mathcal{L}$.}%
\end{figure}
The water-substrate interface is assumed to carry a constant surface charge density~$\rhomitdimen_{\!s}$ corresponding to a dielectric substrate, whereas charges are supposed to be absent at both the oil-substrate and the oil-water interface. All interfaces are assumed to be infinitely thin and uncurved, constituting a wedge geometry for the water phase. We therefore investigate the system on a length scale large enough to neglect the effects of long-range van~der~Waals forces~\cite{deGennes1985,Brochard-Wyart1991,Bonn2009}, but small enough to neglect the interfacial curvature due to the electric field~\cite{Chou2001} or gravitational forces. In equilibrium, the electrostatic potential distribution~$\phimitdimen$ within the water phase (assumed to be a 1-1 electrolyte), can be modeled by the Poisson-Boltzmann equation~$\mitdimen{\Delta}\phimitdimen = kT/e\lambda_D^2 \sinh\left(e\phimitdimen/kT\right)$, where~$k$ is the Boltzmann constant,~$T$ is the absolute temperature, and~$e$ is the elementary charge. In the case that the electrostatic energy of the ions is much smaller than their thermal energy, implying that~$e\phimitdimen/kT\ll 1$, the Poisson-Boltzmann equation can be linearized to yield
\begin{equation}
\mitdimen{\Delta}\phimitdimen=\frac{1}{\lambda_D^2}\phimitdimen
\label{eq:modhelm}
\end{equation}
which is commonly referred to as the Debye-H\"uckel approximation. For general values of the dielectric permittivities~$\varepsilon_w$, $\varepsilon_{np}$ and~$\varepsilon_s$, the potential in all of the three phases would have to be calculated because of the coupling through the interfacial conditions. The corresponding conditions for the normal components of the electric field can be written as~$\vec{n}\cdot(\varepsilon_{w}\mitdimen{\nabla}\phimitdimen_{w}-\varepsilon_{i}\mitdimen{\nabla}\phimitdimen_{i})=\rhomitdimen_{i}$, where~$i$ stands for~$s$ or~$np$, $\vec{n}$ is the normal vector pointing out of the water phase and~$\rhomitdimen_{i}$ the surface charge density at the respective boundaries~\cite{Jackson1998}, with~$\rhomitdimen_{np}=0$. If we assume that~$\varepsilon_i/\varepsilon_w\ll1,~i\in[s,np]$ corresponding to the comparatively high permanent dipole moment of water molecules, the interfacial conditions are simplified to~$\smash{\vec{n}\cdot\mitdimen{\nabla}\phimitdimen_{w}} \approx \rhomitdimen_{i}/\varepsilon_{w}$ (again~$i\in[s,np]$; cf. \bild{fig:Physical_problem}). This simplification allows for a decoupling of the electrostatic potential in the water phase from the potentials in the remaining phases, so that we may focus solely on the water phase. At this point, we introduce nondimensional quantities by scaling the potential~$\phimitdimen$ with~$\rhomitdimen_{s}\lambda_D/\varepsilon_w$, and the coordinates~$\mitdimenflach{x}$, $\mitdimenflach{y}$ and~$\mitdimenflach{r}$ with the Debye length~$\lambda_D$, respectively ([cf.~\bild{fig:Physical_problem}]. Then \gleich{eq:modhelm} and the boundary conditions for the problem transform into
\begin{align}
 & \ohnedimen{\Delta}  \phiohnedimen =\phiohnedimen\label{eq:dimmodhelm} \\
 & \hphantom{\nabla}\!\!\left.\phiohnedimen\right|_{\ohnedimenflachb{y}\to\infty} =0\label{eq:rby}\\ 
 & \ohnedimen{\nabla}  \!\!\left.\phiohnedimen\!\cdot\!\vec{e}_{\ohnedimenflachb{x}}\right|_{\ohnedimenflachb{x}\to\infty} =0\label{eq:rbx}\\
 & \ohnedimen{\nabla} \!\!\left.\phiohnedimen\!\cdot\!\vec{e}_\theta\right|_{\theta=0} =-1\label{eq:rb0}\\
 & \ohnedimen{\nabla}  \!\!\left.\phiohnedimen\!\cdot\!\vec{e}_\theta\right|_{\theta=\alpha+\pi/2} =0\label{eq:rbalpha}
\end{align}
\gleich{eq:dimmodhelm} together with the boundary conditions~\eqref{eq:rby}--\eqref{eq:rbalpha} can be solved analytically using the Kontorovich-Lebedev transform~\cite{Fowkes1998,Kang2003a,Kang2003b,Chou2001,Monroe2006a,Yakubovich1996}. This method provides representations of the solution in the form of one-dimensional integrals that are usually evaluable only by means of costly numerical quadrature. In general, such integral representations do not reveal the functional dependence of the solution on physical parameters, for example, allowing for the study of the decay behavior or the uncovering of characteristic length scales. Only an asymptotic evaluation of the integrals is possible~\cite{Fowkes1998,Chou2001}, yielding approximate solutions of limited validity.\par
In this study, we develop a semi heuristic approximation of the electrostatic potential distribution described by \gleich{eq:modhelm} in a three-phase contact region. The paper is organized as follows. First an approximation for the shape of the equipotential lines is developed and compared with numerical results. From this model, a characteristic length scale over which the oil-water interface influences the double layer structure is extracted. Since the derived expression is implicit with respect to the potential, the latter needs to be evaluated by means of Newton's algorithm using a procedure described at the end of the paper.\par
Before developing the approximation, it is instructive to analyze the general properties of the solution as they can be readily extracted from the boundary conditions~\eqref{eq:rby}--\eqref{eq:rbalpha}. Condition~\eqref{eq:rby} forces the potential to vanish at~$\ohnedimenflach{y}\to\infty$, whereas~\eqref{eq:rb0} determines the normal electric field by prescribing the surface charge density. Condition~\eqref{eq:rbalpha} implies that the electric field vector is parallel to the oil-water interface for~$\theta=\alpha+\pi/2$, whereas condition~\eqref{eq:rbx} expresses the fact the electric field becomes normal to the substrate surface for~$\ohnedimenflach{x}\to\infty$, approaching asymptotically the far field solution~$\phiohnedimen(\ohnedimenflach{x}\to\infty,\ohnedimenflach{y})=\exp(-\ohnedimenflach{y})$. In addition, the Green's function of \gleich{eq:modhelm} has a dominant exponential decay characteristic~\cite{Jackson1998}. Therefore, regarding the influence of the contact line region as a perturbation to the electric double layer extending into the half-space with positive values of~$x$, it is reasonable to assume an exponential transition between the potential in the far-field and close to the contact line that occurs over a characteristic length scale describing the range of the boundary influence. This behavior can be captured directly by approximating the shape of the equipotential lines through a function~$\ohnedimenflach{y}=f(\ohnedimenflach{x})$. In order to parametrize~$f(\ohnedimenflach{x})$, we introduce a pair of coordinates~$(\ohnedimenflach{\eta}_{0},\ohnedimenflach{y}_{0})$ as depicted in \bild{fig:Variables}, where~$\ohnedimenflach{y}_{0}$ is measured at~$x\to\infty$. Note that~$f(\ohnedimenflach{x})$ also depends on~$\alpha$ as well as on~$\ohnedimenflach{y}_{0}$ though we do not explicitly include this dependence in the notation for brevity. For the function~$f(\ohnedimenflach{x})$ we choose the ansatz
\begin{equation}
f(\ohnedimenflach{x})=A-B\hspace{1pt}\euler{-C\ohnedimenflach{x}}
\label{eq:ansatz}
\end{equation}
subject to the conditions
\begin{align}
\left.\partiell{f}{\ohnedimenflach{x}}\right|_{\ohnedimenflachb{x}=-\ohnedimenflachb{\eta}_{0}\sin\alpha}& = \tan\alpha\label{eq:rbfa} \\
f(\ohnedimenflach{x}\to\infty) & =\ohnedimenflach{y}_{0}\label{eq:rbfb}\\
f(\ohnedimenflachb{x}=-\ohnedimenflach{\eta}_{0}\sin\alpha)&=\ohnedimenflach{\eta}_{0}\cos\alpha\label{eq:rbfc}
\end{align}
where condition~\eqref{eq:rbfa} corresponds to~\eqref{eq:rbalpha} and condition~\eqref{eq:rbfb} corresponds to~\eqref{eq:rbx}, respectively. Combining the ansatz~\eqref{eq:ansatz} with the conditions~\eqref{eq:rbfa}--\eqref{eq:rbfc} yields
\begin{equation}
f(\ohnedimenflach{x})=\ohnedimenflach{y}_{0}-\left(\ohnedimenflach{y}_{0}-\ohnedimenflach{\eta}_{0}\cos\alpha\right)\euler{-\tan\alpha\frac{\ohnedimenflachc{x}+\ohnedimenflachc{\eta}_{0}\sin\alpha}{\ohnedimenflachc{y}_{0}-\ohnedimenflachc{\eta}_{0}\cos\alpha}}
\label{eq:fvonx}
\end{equation}
Since the choice of the potential value completely determines~$\ohnedimenflach{y}_{0}$ for each equipotential line through the far-field relation
\begin{equation}
\phiohnedimen=\euler{-\ohnedimenflach{y}_{0}}
\label{eq:farfield}
\end{equation}
it only remains to find a relation between~$\ohnedimenflach{y}_{0}$ and~$\ohnedimenflach{\eta}_{0}$ in order to eliminate~$\ohnedimenflach{\eta}_{0}$ from \gleich{eq:fvonx}. For this purpose we utilize a numerical solution of \gleich{eq:dimmodhelm} subject to the boundary conditions~\eqref{eq:rby}--\eqref{eq:rbalpha} obtained by means of the commercial finite element solver COMSOL Multiphysics\textsuperscript{\textregistered}. The parameters of the numerical calculations, especially the grid structure and size, were chosen such that grid-independent results were obtained. \bild{fig:Zusammenhang} shows the results for different values of~$\alpha$.
\begin{figure}[ht]%
\centering%
\includegraphics{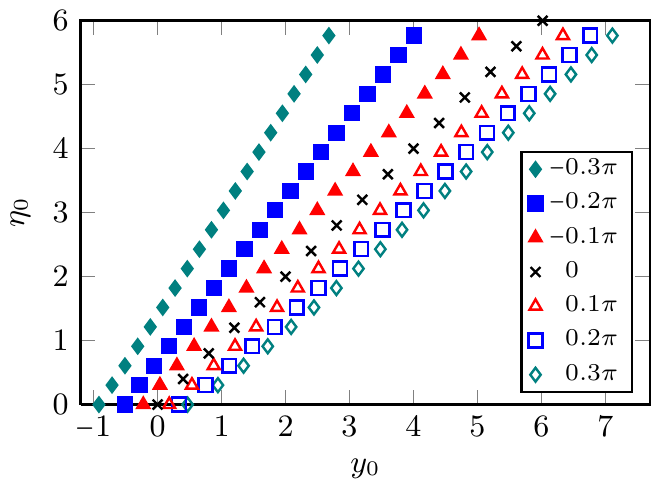}%
\caption{Relation between the coordinates~$\ohnedimenflach{y}_{0}$ and~$\ohnedimenflach{\eta}_{0}$ for different values of $\alpha$ extracted from numerical results}%
\label{fig:Zusammenhang}%
\end{figure}
Obviously, a linear relation~$\ohnedimenflach{y}_{0}\propto\ohnedimenflach{\eta}_{0}$ may serve as an excellent approximation for~$-0.2\pi\leq\alpha\leq0.2\pi$, whereas linearity is slightly violated for~$|\alpha|>0.2\pi$. Therefore we assume that
\begin{equation}
\ohnedimenflach{y}_{0}=c\hspace{1pt}\ohnedimenflach{\eta}_{0}+d
\label{eq:linear}
\end{equation}
where the dependence of~$c$ and~$d$ on~$\alpha$ is suppressed in the notation. \gleich{eq:linear} in conjunction with \gleich{eq:farfield} implies that the potential distribution at the oil-water interface~$\phiohnedimen_{ow}$ can be approximated by the function
\begin{equation}
\phiohnedimen_{ow}=\euler{-c\hspace{1pt}\ohnedimenflach{\eta}_{0}-d}\definiertb b\hspace{1pt}\euler{-c\hspace{1pt}\ohnedimenflach{\eta}_{0}}
\label{eq:phiow}
\end{equation}
In the next step, we propose expressions for the coefficients~$c$ and~$d$~(or~$b$) from analytical considerations and compare them to numerical results. The relation for~$b$ is found from an asymptotic evaluation of the integral representation of the potential~$\phiohnedimen$ at the point~$(\ohnedimen{x},\ohnedimen{y})=(0,0)$ to be of the form~\cite{Chou2001,Kang2003a}
\begin{equation}
\phiohnedimen_{ow}(\ohnedimenflach{\eta}_{0}=0)=\phiohnedimen(\ohnedimen{x}=0,\ohnedimen{y}=0)=b=\mathrm{e}^{-d}=\frac{\pi}{\pi+2\alpha}
\label{eq:banalytisch}
\end{equation}
Furthermore, the gradient of~$\phiohnedimen$ at~$(\ohnedimen{x},\ohnedimen{y})=(0,0)$ is parallel to the oil-water interface [cf. \bild{fig:Variables}] because of the boundary condition~\eqref{eq:rbalpha}, implying that its absolute value is given by~$|\partial\phiohnedimen_{ow}/\partial\ohnedimenflach{\eta}_0|$. In addition, the gradient has to fulfill condition~\eqref{eq:rb0}. Therefore it follows from a comparison between the boundary conditions for~$\phiohnedimen$ at~$\ohnedimenflach{x}\to0$ and~$\infty$ that
\begin{equation}
\frac{1}{\cos\alpha}\left.\partiell{\phiohnedimen}{\ohnedimenflach{y}_0}\right|_{\ohnedimenflach{x}\to\infty,~\ohnedimenflach{y}_0=0}=\left.\partiell{\phiohnedimen_{ow}}{\ohnedimenflach{\eta}_0}\right|_{\ohnedimenflach{x}=0,~\ohnedimenflach{\eta}_0=0}\quad\Longrightarrow\quad c=\frac{1}{b\cos\alpha}=\frac{\pi+2\alpha}{\pi\cos\alpha}
\label{eq:canalytisch}
\end{equation}
In order to compare the expressions~\eqref{eq:banalytisch} and~\eqref{eq:canalytisch} with the numerical solution, we fit \gleich{eq:phiow} to the numerically calculated potential distribution at the oil-water interface and plot the resulting coefficients~$b$ and~$c$, as well as their proposed dependencies on~$\alpha$, in \bild{fig:vergleichbc}.
\begin{figure}[ht]%
\centering%
\subfloat[]{
\begin{minipage}[b][4.5cm][b]{0.45\textwidth}
\includegraphics{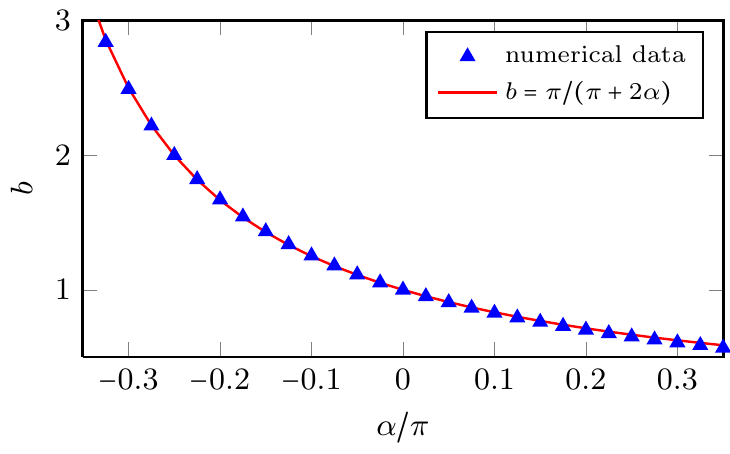}\label{fig:bvergleichbc}
\end{minipage}
}
\subfloat[]{
\begin{minipage}[b][4.5cm][b]{0.45\textwidth}
\includegraphics{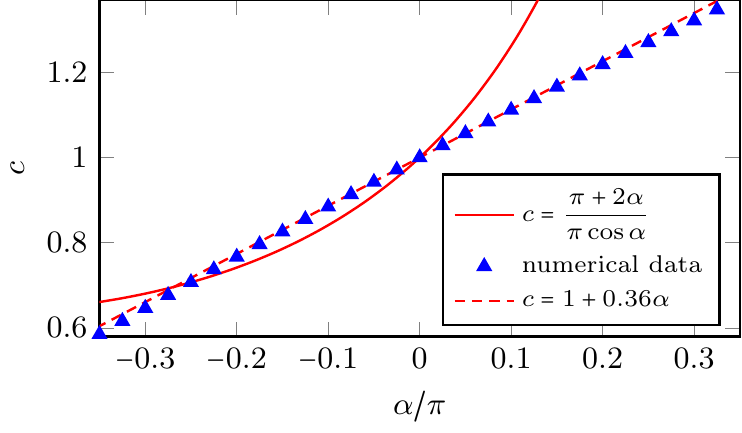}\label{fig:cvergleichbc}
\end{minipage}
}
\caption{ Comparison of the forms~\eqref{eq:banalytisch} and~\eqref{eq:canalytisch} for~$b$ and~$c$ with coefficients found from fitting \gleich{eq:phiow} to the numerical solution}%
\label{fig:vergleichbc}%
\end{figure}
While \gleich{eq:banalytisch} predicts the values accurately [\bild{fig:bvergleichbc}], expression~\eqref{eq:canalytisch} turns out to be inappropriate for modeling the coefficient~$c$ [\bild{fig:cvergleichbc}], which can be reproduced by the linear fit~$c=1+0.36\alpha$ instead. A careful investigation reveals that, although \gleich{eq:canalytisch} correctly predicts the gradient of~$\phiohnedimen$ at~$(\ohnedimen{x},\ohnedimen{y})=(0,0)$, the intention to construct an accurate approximation to the entire potential distribution at the oil-water interface rather demands a linear relation between~$c$ and~$\alpha$. This contradiction is caused by the fact that, except for~$\alpha=0$, the potential distribution at the oil-water interface deviates from an exponential function. As a result, the function~$f$ modeling the shape of the equipotential lines is given by
\begin{equation}
\begin{split}
f(\ohnedimenflach{x})=\ohnedimenflach{y}_{0}-\left(\ohnedimenflach{y}_{0}-\frac{\ohnedimenflach{y}_{0}-d}{c}\cos\alpha\right)\euler{-\tan\alpha\frac{c\ohnedimenflachc{x}+(\ohnedimenflach{y}_{0}-d)\sin\alpha}{c\ohnedimenflachc{y}_{0}-(\ohnedimenflach{y}_{0}-d)\cos\alpha}}\\
\text{with}~d=\ln\left(\frac{\pi+2\alpha}{\pi}\right)\,;~c=1+0.36\,\alpha\,;~\ohnedimenflachc{y}_{0}=-\ln\phiohnedimen
\end{split}
\label{eq:finalfvonx}
\end{equation}
Since up to this point we have only made sure that the model reproduces the potential at the oil-water interface, it is necessary to compare the equipotential lines in the full domain to the numerical results. Examples are depicted in \bild{fig:Potlinien} for four different values of~$\alpha$.
\begin{figure}[ht]%
\centering
\subfloat[$\alpha=-0.25\pi/2$]{
\includegraphics{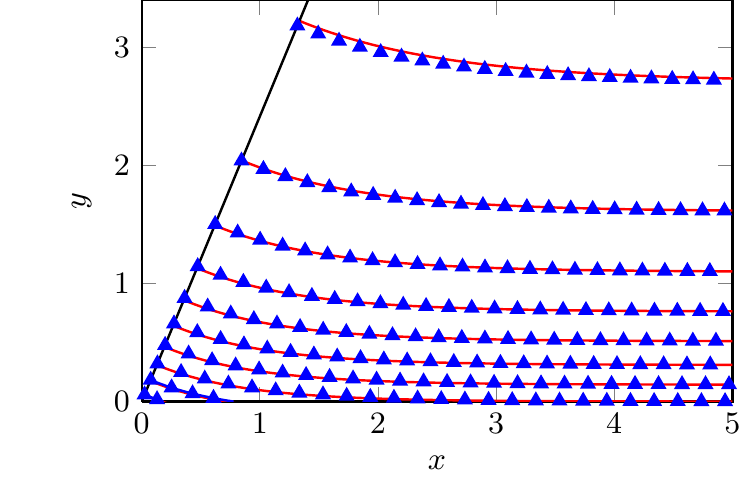}\label{fig:-25Potlinien}
}
\subfloat[$\alpha=-0.1\pi/2$]{\includegraphics{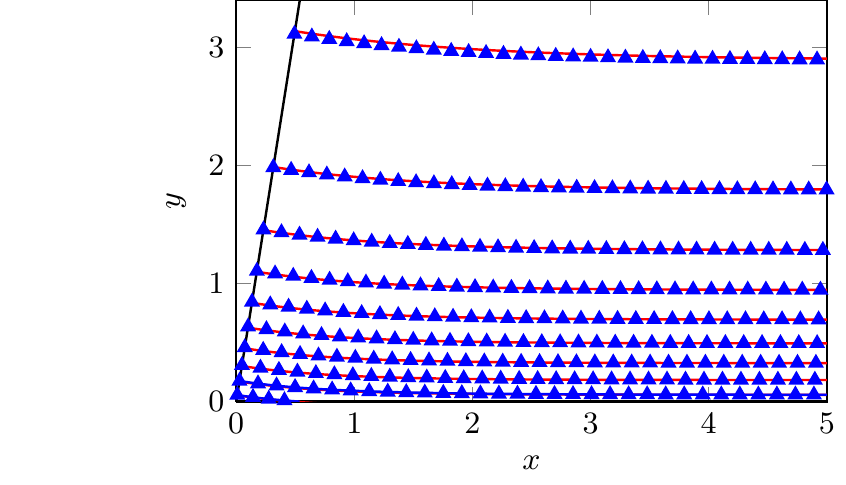}\label{fig:-01Potlinien}}

\subfloat[$\alpha=0.1\pi/2$]{\includegraphics{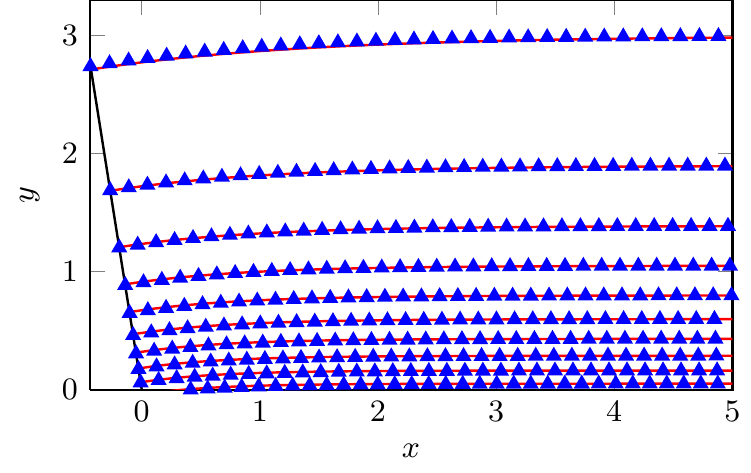}\label{fig:01Potlinien}}
\subfloat[$\alpha=0.25\pi/2$]{\includegraphics{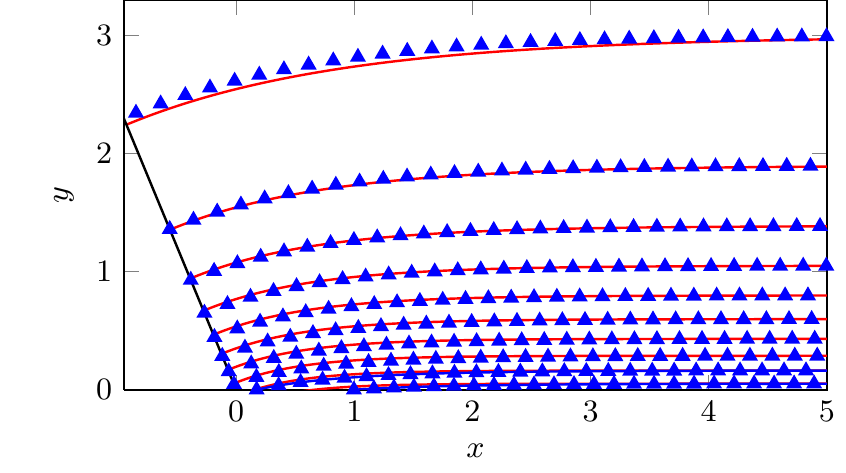}\label{fig:25Potlinien}}
\caption{ Comparison of the equipotential lines (solid lines) calculated from \gleich{eq:finalfvonx} to the numerical results (symbols) for several values of~$\alpha$; lines and data points, from bottom to top, correspond to $\phi=\pi(10/11,\,9/11,\dotsc,\,1/11)/(\pi+2\alpha)$ for~$\alpha<0$ and to $\phi=(10/11,\,9/11,\dotsc,\,1/11)$ for~$\alpha>0$, respectively}%
\label{fig:Potlinien}%
\end{figure}
The agreement within the range~$|\alpha|\lesssim0.25\pi/2$ is excellent, confirming that the shape of the equipotential lines in this range can be approximated by exponential functions. For larger angles~$0.25\pi/2\lesssim|\alpha|\lesssim0.4\pi/2$ (not shown in this paper), \gleich{eq:finalfvonx} still provides a good approximation to the equipotential lines exhibiting only small deviations. Beyond this region, the agreement between model and numerical results deteriorates with increasing values of~$|\alpha|$, but at least qualitative agreement can be observed for~$0.4\pi/2\lesssim|\alpha|\lesssim0.5\pi/2$. Note that the linear behavior~$\ohnedimenflach{y}_{0}\propto\ohnedimenflach{\eta}_{0}$ shown in \bild{fig:Zusammenhang} holds for values up to~$|\alpha|\lesssim0.4\pi/2$ at least.\par
As a main result, from Eqs.~\eqref{eq:fvonx} and~\eqref{eq:finalfvonx} we can extract a characteristic length scale that is given by [cf.~\bild{fig:Variables}]
\begin{equation}
\mathcal{L}=\frac{\ohnedimenflach{y}_0-\ohnedimenflach{\eta}_0\cos\alpha}{\tan\alpha}\lambda_D=\frac{\ohnedimenflach{y}_{0}\left(c-\cos\alpha\right)+d\cos\alpha}{c\tan\alpha}\lambda_D
\label{eq:L}
\end{equation}
The length scale~$\mathcal{L}$ in \gleich{eq:L} is proportional to the Debye length~$\lambda_D$, but is also dependent on the angle~$\alpha$ and the far-field coordinate~$\ohnedimenflach{y}_{0}$, and describes how far the influence of the contact-line region extends into the water phase until the electric double layer finally reaches its far-field equilibrium structure. For small values of~$\alpha$ we find
\begin{equation}
\frac{\mathcal{L}}{\lambda_D}=\frac{2}{\pi}+0.36\, \ohnedimenflach{y}_{0} + \left[-\frac{2}{\pi^2}\left(1+0.36\,\pi\right)+\left(\frac{1}{2}-0.36^2\right) \ohnedimenflach{y}_{0}\right]\alpha+\mathcal{O}(\alpha^2)
\label{eq:Lentw}
\end{equation}
Expansion~\eqref{eq:Lentw} reveals an important property of the potential field, namely a finite value of the length scale~$\mathcal{L}$ appearing even for~$\alpha\to0$. While for vanishing~$\alpha$ the prefactor of the exponential term in \gleich{eq:finalfvonx} also vanishes and thus cancels the effect of~$\mathcal{L}$, there is always an exponential decay over a finite length scale for arbitrarily small~$\alpha\not=0$.\par
Besides the characteristic length scale~$\mathcal{L}$ in \gleich{eq:L}, the approximation~\eqref{eq:finalfvonx} also yields an implicit expression of the form~$\ohnedimenflach{y}=f(\ohnedimenflach{x},\phiohnedimen)$, which cannot be solved for~$\phiohnedimen$ analytically. Therefore, if the potential at a certain point shall be evaluated, numerical methods are needed. Note that in contrast to the complex numerical evaluation of an integral representation of the analytical solution~\cite{Kang2003a}, the numerical method to solve~\gleich{eq:finalfvonx} may be the inexpensive Newton algorithm for nonlinear equations~\cite{Press2007}. The evaluation of the potential~$\phiohnedimen$ at a given point~$(x,y)$ requires the following steps. Therein, we exploit the much smaller variation of the field~$\ohnedimenflach{y}_0(\ohnedimenflach{x},\ohnedimenflach{y})$ with the coordinate~$\ohnedimenflach{y}$ compared to~$\phiohnedimen(\ohnedimenflach{x},\ohnedimenflach{y})$, whose exponential variation is removed by \gleich{eq:farfield}.
\begin{enumerate}
	\item Estimate a starting value for~$\ohnedimenflach{y}_0$ using the far-field relation~$\ohnedimenflach{y}_0=\ohnedimenflach{y}~\text{as}~\ohnedimenflach{x}\to\infty$
	\item Solve \gleich{eq:finalfvonx} for~$\ohnedimenflach{y}_0$ iteratively by means of Newton's algorithm which yields fast convergence within only a few iterations
	\item Calculate~$\phiohnedimen(\ohnedimenflach{x},\ohnedimenflach{y})$ from \gleich{eq:farfield}
\end{enumerate}
In this way, the model~\eqref{eq:finalfvonx} not only provides characteristic information on the double layer structure but also allows for an inexpensive calculation of the complete potential distribution. Note that by using the nonlinear Poisson-Boltzmann equation instead of the Debye-H\"uckel approximation a similar model can be obtained, though the value of the electrostatic potential on the three-phase contact line is not known in this case.

%

\end{document}